# A new class of photocatalytic materials and a novel principle for efficient water splitting under infrared and visible light: MgB$_2$ as unexpected example

V. G. Kravets and A. N. Grigorenko

*University of Manchester, School of Physics and Astronomy, Manchester, United Kingdom*
*vasyl.kravets@manchester.ac.uk*

**Abstract:** Water splitting is unanimously recognized as environment friendly, potentially low cost and renewable energy solution based on the future hydrogen economy. Especially appealing is photocatalytic water splitting whereby a suitably chosen catalyst dramatically improves efficiency of the hydrogen production driven by direct sunlight and allows it to happen even at zero driving potential. Here, we suggest a new class of stable photocatalysts and the corresponding principle for catalytic water splitting in which infrared and visible light play the main role in producing the photocurrent and hydrogen. The new class of catalysts – ionic or covalent binary metals with layered graphite-like structures – effectively absorb visible and infrared light facilitating the reaction of water splitting, suppress the inverse reaction of ion recombination by separating ions due to internal electric fields existing near alternating layers, provide the sites for ion trapping of both polarities, and finally deliver the electrons and holes required to generate hydrogen and oxygen gases. As an example, we demonstrate conversion efficiency of ~27% at bias voltage $V_{bias}$=0.5V for magnesium diboride working as a catalyst for photoinduced water splitting. We discuss its advantages over some existing materials and propose the underlying mechanism of photocatalytic water splitting by binary layered metals.

___________________________________________________________________________________

## 1. Introduction

The decomposition of water by sunlight has long been recognized as a potentially important reaction to harvest and store solar energy for two main reasons. First, the process of water splitting (being a thermodynamically uphill reaction) can be powered by more than half of the photons in the solar spectrum, providing an effective method for solar energy storage [1-3]. Second, the water splitting and photocurrent generation produces a clean energy carrier, hydrogen, which regenerates water without adversely altering the environment.

As water is largely transparent to most of the solar spectrum, the key challenge in carrying out the reaction efficiently and, more important, inexpensively is to find a suitable catalyst that can absorb sunlight broadly, transfer the energy to the ions, strongly suppress the inverse reaction of ion recombination and hence catalyse water splitting with high specificity. Solar energy is distributed in ultraviolet, visible, and infrared light with a proportion of about 7%:50%:43%. However, traditional photocatalysts based on metal oxides [2,3], sulphides [4,5], nitrides [6], graphene-semiconductor nanocomposites [7,8], etc., are mostly active only under ultraviolet irradiation, while other photocatalysts which absorb visible light are not stable during the reaction process. For example, for metal sulphide photocatalysts show strong photocorrosion and hence provide relatively low quantum yield under visible light (usually under 10%), leading to inefficient usage of sunlight. During 2014 there have been multiple demonstration of organohalide-perovskite solar cells with efficiencies > 12%. However, the best perovskite candidates are based on lead ($CH_3NH_3PbI_{3-x}Cl_x$ with $0<x<3$) which are very toxic (contains PbI) and their instability limits the lifetime [9,10].

To address these problems, various methods have been pursued. For example, the band gap of a photocatalyst was reduced to the visible light through band-gap engineering such as cation or anion (co-)doping to the semiconductor or combining two semiconductors with larger and smaller bandgaps [8,11]. The photocorrosion was reduced by employing metal-free photocatalysts [12,13]. However, it is safe to say that the abundant, stable and effective photocatalyst for water splitting is still lacking and the water splitting technology does not utilise the whole range of solar light capable of producing the reaction [2,14]. To break through these restrictions, a new mechanism and materials are needed. An optimal material should combine an ability to dissociate the water molecules, suppress the inverse ion recombination reaction and utilize light in the infrared (IR) and visible (VS) range and to remain stable in contact with water.

Generally, the photocatalytic water splitting makes use of semiconducting, dielectric and so-called metal free materials which demonstrate the mentioned shortcomings. (In parenthesis we note that recently there are some reports on using metals in various complex compounds.) We try to go in a different direction: our photocalytic method is based on earth abundant metals which possess a very high concentration of conduction electrons. These materials can absorb light from broad spectral region (IR and VS) and a huge number of electrons take part in photocatalysts processes resulting in photocurrent enhancement. It is worth noting that the water splitting requires energy of 1.23eV. Hence, the visible and infrared light (with the wavelength from 650nm to 1000nm) can in principle induce water splitting reaction. At the same time the energy difference between the photon energy and 1.23 eV is lost for energy storage in the form of hydrogen. This implies that each photon capable of producing water splitting generates exactly the same amount of hydrogen and hence stores the same energy. Therefore, the part of the solar spectrum with the maximal number of the photons will provide the maximal yield for energy storage. It is easy to find that the wavelength with maximal number of solar photons (as a function of wavelength) corresponds to 640nm and lies at the edge of VS light very close to IR. As a result, the catalyst working in VS till near IR are important for hydrogen energy storage using water splitting.

Here we present a direct experimental evidence for photogeneration of the electric current from dissociated water molecules using a new class of catalytic materials – ionic/covalent binary metals with layered graphite-like structures (or, simply, binary layered metals (BLM)). BLM consists of alternative layers of metals (Al, Mg, Cr, etc.) and metalloids (B) where the ionic (or covalent) exchange between layers results in alternating charges (electrons and holes) sitting on alternating layers. These charges lead to three important properties which facilitate the reaction of water splitting. First, the presence of the electron and hole plasmas provides strong interaction with light of IR-VS spectra by BLM and an effective transfer of solar energy to ions. Second, alternating charges of the layers separate ions spatially by providing strong electric fields (in analogy to an array of *p-n* junctions) and trapping sites for ions of both polarities which effectively suppress the inverse reaction of ion recombination. Third, the overall metallic properties allow BLM to deliver the electrons and holes required to generate hydrogen and oxygen gases and thus complete the process of water splitting. As an example, we discuss in detail the water splitting by overall metallic $MgB_2$ under illumination by IR-VS light spectra. We show that the electrodes produced from $MgB_2$ by a simple spraying process can demonstrate photocatalytic water splitting with high solar energy conversion efficiency. We found that a $MgB_2$ electrode can replace the expensive Pt counter electrode in photochemical cells and two $MgB_2$ layers can act as both photoanode and cathode yielding a conversion efficiency of ~27% at small bias voltage $V_{bias}$=0.5 V. Our results pave the way for engineered the optimal materials for solar water splitting based on compounds with typical graphite-like structures, such as $AlB_2$, $NbB_2$, $MoB_2$, $TaB_2$, $TiB_2$, $HfB_2$, and $CrB_2$. We also discuss a novel reaction mechanism for water splitting by BLM. Such visible-infrared-light-driven water splitting opens a new future for solar energy conversion.

2. **Sample preparation and characterization**

The field of photoelectrochemistry emerged as a competitive energy conversion technology as a result of the work of Fujishima and Honda [2] on water splitting near a UV-irradiated $TiO_2$ electrode. Following this pioneering work, we built a photoelectrochemical cell schematically shown in Fig. 1(a). The $MgB_2$-base electrode fabricated on the top of different noble-metal films was placed in a photoelectrochemical cell (Fig. 1(a)) filled by $H_2O$ or $D_2O$. The spacing between the two electrodes (normally $MgB_2$ and Pt (Pt electrode has diameter 1.5mm and length 25mm)) was about 10 mm. The photocurrent response was measured using a digital voltmeter Keithley 2400 in the current mode. The photocatalytic cell was a cylindrical glass vessel with height of 5 cm and diameter of 2 cm. The area of the light irradiation was approximately 1 $cm^2$. We used a VS-IR light source with total intensity of ~15 $mW/cm^2$ (which is one order of magnitude smaller than the total solar irradiance) with the peak spectral intensity achieved at 700 nm. A source with almost no ultraviolet (UV) radiation was deliberately chosen in order to remove the possibility of direct excitation of intraband transitions from deep electronic levels to empty states in the vicinity of the Fermi level. We also performed gas chromatography to check the hydrogen production.

A magnesium diboride layer was formed with the help of inexpensive spraying technique from a suspension produced by liquid-phase exfoliation of $MgB_2$ in ethanol solution. In brief, 10 mg of pure $MgB_2$ powder (Sigma Aldrich) was suspended in 50 ml ethanol by sonication at amplitude of 15 W for 10 hours with a homogenizer. The resulting $MgB_2$ flakes were mostly single polycrystals with lateral dimensions of 1-2 μm, as characterized by optical microscopy (Fig. 1(b)). The main advantage of liquid-phase exfoliated suspension is a possibility to prepare large-scale flexible devices on commercial, transparent, and flexible substrates. $MgB_2$ film electrodes were fabricated using inexpensive spray deposition technique. The exfoliated suspension of $MgB_2$ was sprayed onto a clean glass substrates covered by either Au(50nm), or Ag (50nm), or Cu standard foil (100μm) at 120°C using a spray gun.

We started with calibration of our installation by utilising reference electrodes with a $TiO_2$ layer (instead of $MgB_2$ ones). In this case, the photocurrent was generated under illumination by UV LED with the maximum intensity at the wavelength of 380 nm and the half-width of 30 nm. (The photocurrent was not produced under illumination of $TiO_2$ electrode by VS-IR light.) The $TiO_2$ solution in ethanol was prepared in the same way as $MgB_2$ samples. The light intensity of LED was measured by a calibrated spectrometer and the total power was found to be W ~ 1 mW/cm$^2$. The results of the photocurrent measurements are shown in Fig. 2(a). As the light was turned on, the photocurrent increased about 10 times compared to that of the thermal process in the absence of light. This behaviour was fully reversible with the photocurrent responding instantaneously to the introduction and removal of the light flux. It is well known [2,6,8] that excitons appear in a TiO$_2$ anode illuminated by UV light. The photogenerated holes (h$^+$) drift to the contacting water interface where they oxidize H$_2$O (through the reaction 4h$^+$+2H$_2$O→4H$^+$+O$_2$). The electrons traverse the semiconductor layer to the back contact (Au film in our case) and reach the cathode (a Pt electrode, Fig. 1(a)) through the external circuit, where they reduce absorbed protons to form hydrogen according to 2e+2H$^+$→H$_2$ [6,8]. (It is worth stressing again that in case of VS-IR light illumination none of the relevant photoprocesses takes place and electron-hole pairs are not generated in the TiO$_2$.)

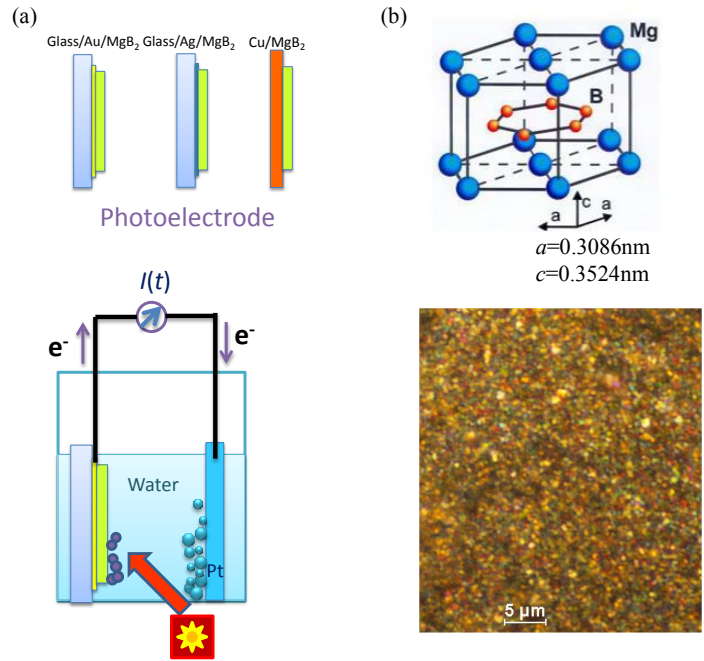

Fig. 1. (a) Schematic design of integrated photoelectrochemical cell using MgB$_2$ as the photoanode and Pt as the cathode in distil water (18.2 MΩ/cm). (b) The optical micro image of a MgB$_2$ layer of about 1 μm in thickness composed of fine grains with a wide distribution of orientations.(Top inset: hexagonal structure of MgB$_2$ consisting of honeycomb B (yellow-red) layers with close-packed Mg (blue) layers between them.)

Figure 2 shows one of the main results of our work – the photocurrent generated by a MgB$_2$ electrode placed into water under illumination by infrared and visible light. The VS-IR-light splitting of water near MgB$_2$ electrodes was measured without applying external voltage. To show the current and electrode stability, we present the photocurrent generation during time of 1-3 hours (Fig. 2(b-f)). Fluctuations of the current with time were caused by bubble formation, perhaps of H$_2$ and O$_2$, on the surface of electrodes, which affects the effective surface area of electrodes. The produced H$_2$ bubbles (clearly seen by eyes) gathered on the cathode surface of Pt electrode, while O$_2$ bubbles were gathered on the MgB$_2$ anode. The conversion efficiency has been calculated using bubble microscopy [15] (and following gas chromatography). These measurements confirmed 0.95% Faraday efficiency (each 2 electrons generated ~0.95 molecules $H_2$). The photon to current conversion ratio is discussed below. Note the overall small decrease in current on the longer time scale, which is mainly due to small deterioration of sprayed $MgB_2$ films and can be eradicated by adding stabilisers. Moreover, when photocurrent-time measurements were carried out in $D_2O$ (Fig. 2(c)), the current was about two times lower than that of obtained in the case of $H_2O$. These results suggest that $MgB_2$-water interface acts as proton donors in this system. We discuss the effectiveness of the process of water splitting under VIS and IR light by magnesium diboride below.

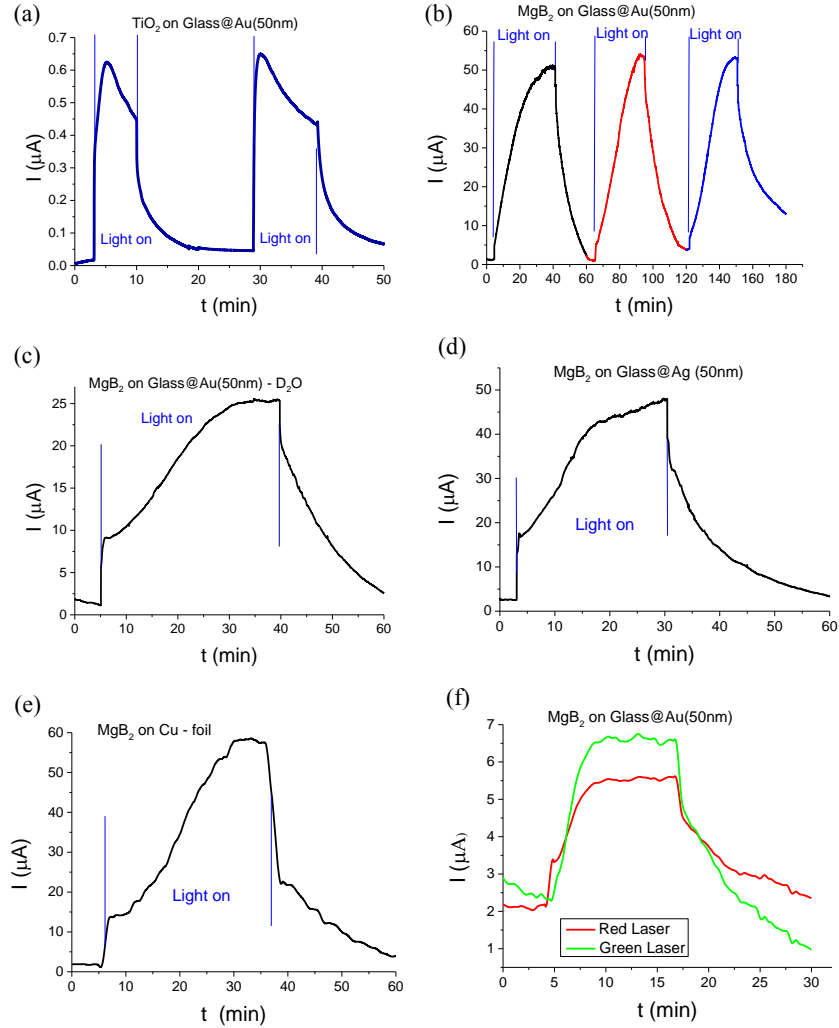

Fig. 2. Photocurrent as a function of time for the integrated water splitting device without external bias voltages under illumination by infrared-visible light: (a) Solar cell based on standard $TiO_2$ photoanode under UV light. (b) Solar cell based on $MgB_2$ layer deposited on Au (50nm) film, $H_2O$ splitting. (c) $MgB_2$ layer deposited on Au (50nm) film, $D_2O$ splitting. (d) $MgB_2$ layer deposited on Ag (50nm) film, $H_2O$ splitting. (e) $MgB_2$ layer deposited on Cu (100 μm) foil, $H_2O$ splitting. (f) $MgB_2$ layer deposited on Au (50nm) film for monochromatic laser irradiations.

The role of the different substrates (glass/Au(50nm), glass/Ag(50nm), and Cu foil (100μm) was investigated. The most efficient electrodes have generated photocurrents of up to ~60 μA cm$^{-2}$ under our VIS-IR light source. The MgB$_2$ photoanodes remained active after 3 hours of testing and MgB$_2$ was found to be stable during the water splitting. Figures 2(d) and 2(e) show that our measurements of the photocurrent were completely reproducible for different devices of approximately the same size. It was observed that photocatalytic properties of MgB$_2$ layers display weak substrate dependence. Using different noble-like metals (Au, Ag, Cu) as an under-layer for the surface MgB$_2$ (Fig. 1(a)), we found that the maximal current was slightly higher on an Au film and a Cu foil rather than on an Ag film. (We also observed slight change in a dark photocurrent, see Figs. 2(b-d)). During the photoreactions, the H$_2$ and O$_2$ gases were produced in the water near MgB$_2$ and Pt electrodes. The hydrogen production was directly related to the photocurrent through the reaction 2H$^+$+2e$^-$=H$_2$.

To check that water splitting reaction is indeed excited by VS-IR range of spectrum we have carried out photocurrent spectroscopy measurements in which the photo-generated current was measured under monochromatic sources (lasers) of various wavelengths. Figure 2(f) shows examples of photocurrents generated by green ($\lambda\approx$540nm) and red ($\lambda\approx$670nm) lasers approximately of the same density power (~5mW/cm$^2$). These measurements suggest that there was no significant spectral decrease of the effectiveness of photocurrent generation toward IR part of the spectrum.

To characterize the MgB$_2$ layers in terms of chemical composition, impurity and disorder before and after the use in the water splitting cell we apply Raman spectroscopy. The Raman spectra were measured using a confocal laser Raman spectrometer (Renishaw) with a 100x microscope. The 514.5 nm line of an Ar+ laser was used for excitation, with the laser power maintained at 1 mW in order to avoid laser heating of the studied materials. Several spots were selected on the same sample to collect the Raman signals in order to make sure that the results were consistent. Figure 3(a) shows the Raman spectra of pure MgB$_2$ powder and MgB$_2$ films measured before and after photogeneration of currents during water splitting. Two broadband features centred at about 590 and 800 cm$^{-1}$ were observed for layered materials. In powder MgB$_2$, the most prominent phonon peak is located at lower frequency (~590 cm$^{-1}$) and assigned to the $E_{2g}$ mode [16-18]. The half-width of the $E_{2g}$ peak in powder MgB$_2$ exceeds 200 cm$^{-1}$, which is attributed to the presence of strong electron-phonon coupling and phonon-phonon interaction [16]. The most notable feature of the $E_{2g}$ mode in the MgB$_2$ layer used for H$_2$O and D$_2$O splitting is slight decrease in both the frequency and the linewidth. The higher frequency Raman band (centred at ~800 cm$^{-1}$ in Fig. 3(a)) has been observed earlier in MgB$_2$ and was attributed to phonon density of states (PDOS) due to disorder [17,18]. It is interesting to note that despite the peak located at about 800 cm$^{-1}$ is not observed for powder MgB$_2$ it becomes the prominent feature in the Raman spectrum for layered samples.

Figure 3(b) shows extinction spectra for layered MgB$_2$ films made on transparent substrate by the spraying technique and liquid solution. (Extinction here is defined as –ln($T/T_0$), where $T$ is the light transmitted through the sample and $T_0$ is the light transmitted through the substrate.) We obtain the same shape of extinction spectra for the fresh and used for water splitting layers which attests stability of our photoelectrodes. The extinction reaches the local minimum at ~600 nm ($\hbar\omega\approx$2.0 eV) and the local maximum at ~1200 nm ($\hbar\omega\approx$1.0 eV). The displayed dependences are in agreement with spectra of optical conductivity [19,20]. For MgB$_2$, the interband contribution to the in-plane optical conductivity $\sigma(\omega)$ contains strong IR peaks with a tail extending to the red part of the visible range and a resonance around 2.6 eV, which is close to the observed value of 2.0 eV [19]. The IR peaks are associated with transitions between two $\sigma$-bands while the peak in the visible range is associated with a transition from $\sigma$ band to the $\pi$ band close to the M point, where a van Hove singularity strongly enhances the density of states [19]. In the IR region the absorption is dominated by metallic-like response, which is characterized by a broad maximum and can be modelled using the Drude expression. Due to contribution from large number of different orientation of $MgB_2$ crystals we cannot see anisotropy in the absorption spectra.

### 3. Results and discussion

We discuss electronic and optical properties BLM using $MgB_2$ as an example. Magnesium diboride becomes superconducting at about 40 K [21]. Boron ions in crystalline $MgB_2$ are packed in honeycomb layers alternating with hexagonal layers of magnesium ions (Fig. 1(b)). The in-plane distance, $b=1.78$ Å, between the boron cations, is less than that between magnesium anions, $a=3.08$ Å. The space separating the boron planes is of the size $c=3.52$ Å. The magnesium ions are positioned above the centers of hexagons formed by boron sites and donate their electrons to the boron planes: $Mg^{2+}[B^-(p^2)]_2$. These $p$ electrons form $\sigma$ and $\pi$ bands and the charged Mg layers shift the electronic energy bands so that the $\pi$ band becomes lower than the $\sigma$ band. The latter crosses the Fermi level, providing the light- and heavy-hole formation. Thus the electronic structure is formed by two characteristic charge carriers in $MgB_2$: "heavy" holes of B $\sigma$ bands and "light" electrons and holes of B $\pi$ bands [22,23]. Moreover, theory predicts a sharp collective mode whose origin is the strong coherent charge fluctuation between parallel sheets of B and Mg atoms [24].

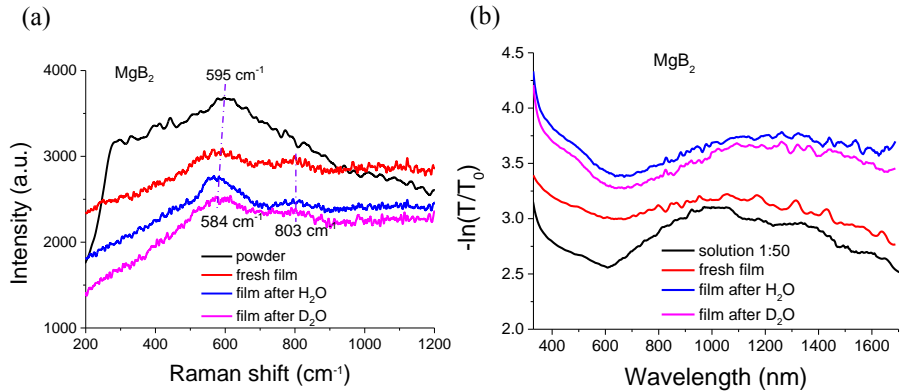

Fig. 3. Optical spectra of samples. (a) The Raman spectra of pure powder $MgB_2$ and layered $MgB_2$ samples freshly prepared and after treatment in photogenerated water split cell by $H_2O$ and $D_2O$, respectively (b) The extinction spectra of layered $MgB_2$ freshly prepared and after treatment in photogenerated water split cell by $H_2O$ and $D_2O$ (For comparison, spectrum of liquid solution of $MgB_2$ powder in ethanol (solvent in proportional 1:50) is given).

Magnesium diboride (and other BLM) possesses interesting plasmon collective excitation modes which are observed in reflectivity. Ellipsometric measurements [19] demonstrate that the conductivity and the plasma frequency are both tensors with two different components: one parallel to the $aa$-plane and another along the $c$-axis (Fig. 1(b)). Note that the plasma edge parallel to the $aa$-plane is at about $\omega_{p,a} \sim 2$ eV while the one along the $c$-axis is at $\omega_{p,c} \sim 2.7$ eV [19]. A nice manifestation of the plasma edge anisotropy is the multicolour appearance of magnesium diboride seen in the optical microscopy image (Fig. 1(b)). Our sample spectacularly changes colour from a blueish silver for $E$ parallel to B planes $aa$ to the yellow for $E$ along $c$ axis (Fig. 1(b)). The explanation comes naturally from the reflectivity dependences: for the $c$-axis polarization, it is a sharp plasma edge at about 2.5 eV, the same as of gold, which makes the sample yellow. For the *aa*-plane polarization, the plasma edge is smeared due to the strong interband transition at 2.8 eV which makes the colour blue silver. At the same time, the measured absorption spectra of our films demonstrate a broad plasma edge at about $\hbar\omega \approx 2$ eV which is most probably connected with orientation disorder of $MgB_2$ flakes and the size effect. Data presented for different samples at Fig. 3(b) do not reveal any significant influence of the sample fabrication procedures on the optical behaviour in the VS-IR region. We found that the dynamics of the photocurrent produced during the water splitting is dissimilar for electrodes made of $MgB_2$ as compared to reference $TiO_2$. In case of a $TiO_2$ photoanode (Fig. 2(a)), the current quickly saturates with time and then drops. For $MgB_2$, the scenario is different: the photocurrent jumps at the first stage and then monotonically increases during the longer second stage (Fig. 2(b-e)). The photogeneration of current in $MgB_2$-based solar cell is unexpectedly efficient, see below, suggesting that a new model may be needed to describe water splitting mechanism in such systems.

Here we propose a new photosynthesis scenario inspired by a two-band electronic structure of BLM. It is recognized [14] that an effective photo-catalytic reaction of water splitting includes several important processes: i) absorption of photons to form electron-hole pairs driving creation of $H^+$ and $OH^-$ ions, ii) suppression of inverse reaction of ion recombination by spatial charge separation and iii) construction of surface reaction sites for $H_2$ and $O_2$ evolution. BLM delivers all three possesses required for water splitting.

*3.1 Photon absorption and creation of electron-hole pairs.*
In contrast to semiconductors, where electron-hole pairs are produced by an electron transition from the valence band to the conduction band under (mostly) UV light, the electron-hole pairs in BLM are created by interband transitions between $\pi$ and $\sigma$ bands. This transition has a van-Hove singularity in $MgB_2$ at around 2.0-2.6 eV as discussed above; at the same time, it is quite broad due to a complex zone structure and hence covers the whole visible and IR range. Electron-holes pairs generated deep inside the metal do not survive because of fast and strong Debye screening. However, the surface of metals [25] – the place where the catalytic reaction takes place – provides a nice exception. At the metal surface, the Debye screening is compromised due to electron hole interaction through the surrounding dielectric and transient excitons can exits [25]. The situation is even more favourable for BLM, where alternating layers are charged due to ionic/covalent interaction between layers. This leads to an electron-hole spatial separation (electrons and holes created by incident light could reside in different layers and therefore survive longer) and to dependence of single electron states on the surface termination layer: even the band structure deforms near the surface due to ionic and electron charge densities in alternating layers [26]. For example, in $MgB_2$, Mg layers give electrons to B layers and become positively charged while B layers accept electrons and become negatively charged. In the bulk state, as was shown in [27], the B sheets are over doped by electrons. The Mg ionic charge is able to neutralize the negative charge on the subsurface B planes, giving electrical stability to the material. In other words, Mg atoms donate electrons interstitially and can be considered as an *n*-type doping of the surface states of the clean B-terminated surface, Fig. 4(a). Angle-resolved photoemission spectroscopy confirmed the appearance an additional band assigned to a surface state of the terminated surfaces [28].

It is difficult to evaluate influence of surface effects on the properties of electron states in IBML for our case of randomly distributed crystals with different orientation and sizes. As a rough estimate, we make use of calculations performed for geometry with a flat layer ending in a simple single particle picture. For $MgB_2$, it was found [26] that the B-terminated surface gives a work function of $\Phi_B \approx 5.95$ eV (1.45 eV vs standard hydrogen electrode, SHE, with $\Phi_H \approx 4.5$ eV), while the Mg-terminated surface gives $\Phi_{Mg} \approx 4.25$ eV (-0.25 eV vs SHE). The difference between these energies, $\Phi_B$-$\Phi_{Mg}$, roughly represents the energy required to move electron from B to Mg layer and create an electron-hole pair. Figure 4(b) provides schematic representation of water splitting in $MgB_2$ by surface electronic states. One can see that the top potential of surface electrons is more negative than the redox potential of $H^+/H_2$ (0 eV vs SHE) while the bottom potential of surface electrons is more positive than the redox potential of $O_2/H_2O$ (1.23 eV vs SHE) which is necessary for photo-catalytic water splitting reaction [14,29].

We conclude that unusual surface states of electrons in ionic binary metals are responsible for creation of electron-hole pairs that creates ions of $H^+$ and $OH^-$ in this simple single particle picture. In reality, the situation is more complicated. Indeed, metals interact with light through collective oscillations of electron plasma which yield quantized collective plasmon states. In BLM, plasmons are characterized by plasmon bands which can strongly interact with surface states and affect creation of surface electron-hole pairs [24,30]. The relation between BLM plasmons and surface states is an open question which requires further investigation.

It is important to note that the electron plasma provides negative values of metal permittivity in visible range and hence the light absorption in metals happens in thin (skin) layer where the catalytic reaction takes place. This makes a metal catalyst to be more effective than a semiconductor one as a photon absorbed in the bulk of a semiconductor catalyst is effectively lost for the reaction while the photon reflected from the metal could be used again. Due to the presence of a skin layer and relatively high light reflection from metal, in order to describe effectiveness of the water splitting process, one needs to introduce internal quantum efficiency, $\eta_{IQE}$, which is defined as the number of electron-hole pairs produced per each photon absorbed in the metal [31]. The efficiency $\eta_{IQE}$ can be expressed as [31]: $\eta_{IQE} = \dfrac{I}{P(1-R)} \dfrac{\hbar\overline{\omega}}{e}$, where $I$ is the photo-generated current, $P$ is the light power, $R$ is the reflection coefficient, $\hbar\overline{\omega}$ is the average photon energy and $e$ is the electron charge. To calculate $\eta_{IQE}$ we assume that $I_{ph} \cong 60\mu A$ (Fig. 2) is the maximal photocurrent density for our installation, $\hbar\overline{\omega}$ corresponds to the wavelength of 700nm (1.7eV), $P$=15mW is the power of light source and $R \approx 0.85$ [19] is an "average" reflection coefficient in the spectral region near wavelength of 700nm. This yields the value of $\eta_{IQE} \approx 5$ % for the $MgB_2$ active layer shown in Fig. 2(e). This is a conservative estimate of the quantum efficiency as the geometrical size of our platinum electrode might have been limited the saturation current in our experiments.

*3.2 Suppression of the inverse reaction by charge separation.*

The alternating layered structure of BLM with one layer charged positively and the other negatively ("charged density wave" near the surface) provides an effective way to spatially separate photogenerated ions at the surface of BLM. One can consider the surface of BLM as a set of alternating *p-n* junctions with strong electric fields that perform the charge separation in the same manner as it happens in a *p-n* junction. We can roughly evaluate the electric fields near the BLM surface as $E_s \approx \dfrac{q}{2\varepsilon_w d^2}$, where $q$ is the charge transferred from one layer to another per unit cell, $d$ is an average distance between atoms in the cell and $\varepsilon_w$ is the permittivity of water. For $MgB_2$, we have $q \approx 0.1e$, $d \approx b$ =0.178nm [21] which yield very large electric fields separating ions $E_s \approx 10^8$ V/m. These fields are highly inhomogeneous near the surface of metal where the catalytic reaction takes place, Fig. 4(a). As a result, ions created during water splitting become separated spatially: positive ions $H^+$ travel to the layers with a negative charge (B layers in case of $MgB_2$) while negative ions $OH^-$ travel to the plane with a positive charge (Mg layers in case of $MgB_2$), Figs. 4(a) and 4(b). This leads to strong suppression of the recombination reaction and insures effectiveness of BLM as a photocatalytic converter.

The fact that ions of both polarities can be trapped near BLM surface leads to an important conclusion that BLM can be used as both an anode and a cathode in the catalytic reaction of water splitting. Ordinary, the counter electrode made of Pt is used in splitting cells in order to achieve high efficiency of photoelectrochemical processes. This choice is mostly due to high photo- and electro-catalytic activity and good chemical stability of Pt electrodes. However, being a noble metal, Pt is relatively expensive and hence cannot be straightforwardly applied for large-scale production of water splitting cells. Therefore, a platinum-free efficient cathode would be of great benefit for the cost reduction of water splitting technique. We found that BLM, and $MgB_2$ in particularly, can indeed serve as both a photoanode and a cathode in the water splitting cells and therefore can provide a cost-effective counter electrodes alternative to the noble metal Pt. Figure 4(c) shows the photocurrents generated in a symmetric water splitting cell where both electrodes (the anode and the cathode) are made of $MgB_2$ under the same light source as in Fig. 2(b). We see that the symmetric cell (made of two $MgB_2$ electrodes) generates the photocurrent with high conversion efficiency. The maximal photocurrent is around 6 times larger than that for the case of $MgB_2$ and Pt electrodes shown in Fig. 2(e) under the same illumination. Note that relatively small bias (0.5V) was necessary to break the symmetry of two electrodes in order to generate high values of the photocurrent with the help of cheap $MgB_2$ electrodes. This enhancement of photocurrents can be recalculated in large increase of the efficiency of catalytic water splitting process when two $MgB_2$ electrodes are used as $\eta_{IQE}(V_{bias}=0.5V) \approx 27\%$.

*3.3 Construction of surface reaction sites for $H_2$ and $O_2$ evolution.*
It is well-known that the presence of surface reaction sites for $H_2$ and $O_2$ evolution can strongly enhance the yield of photocatalyst [3,14]. Often, it is the surface nanostructure of catalytic material that provides the reaction sites [32]. This is the case for BLM. Indeed, endings of layers of BLM become the trapping centres for ions of both polarities, see above. The overall metallic nature of BLM, Materials, allows easy transfer of charges (electrons and holes) to the ions and the typical reactions expressed in simplified form as $4H^+ + 4e = 2H_2$ (cathode) and $4h^+ + 2H_2O \rightarrow 4H^+ + O_2$ (anode). This completes the reaction of water splitting.

We can summarise the mechanism of water splitting with the help of BLM as follows: an absorption of a photon creates an electron-hole pair near the metal surface, the pair excites the water splitting reaction and produces corresponding ions, the ions are affected by the electric fields produced by alternating charges on the electrode layers and hence travel to the reaction reduction sites which are the ending of the metallic layers where they recombined with electrons and holes supplied by BLM. Obviously, the described photocatalystic mechanism is not restricted only to water splitting. It can be extended to other photocatalytic process, for example, the splitting of $CO_2$.

For completeness, we confirm that the photocatalytic principles described in this work can be applied not only to BLM based on boron but also to other binary layered metallic structures, e.g. metallic transitional metal dichalcogenides. For example, electrodes with a $NbSe_2$ layer (prepared by the same spraying technique as other electrodes) also effectively absorbs VIS and IR light facilitating the reaction of water splitting and delivers the electrons and holes required to generate hydrogen and oxygen gases. Figure 4(d) plots the photocurrent generated in our water splitting cell with a $NbSe_2$ electrode (and Pt as the cathode) under the same illumination conditions as for Fig. 2. One can see that $NbSe_2$ shows smaller photoelectrocatalytic activity compared to the $MgB_2$ (juxtapose Fig. 2(b) and Fig. 4(d)).

There is only one area where metallic electrodes appear to be in disadvantage as compared to semiconductor ones. Indeed, metal got very large impedance mismatch with water and tend to reflect a lot of light. However, this disadvantage can be easily cured by using plasmonic "black-body" produced by metal-dielectric composites where suitably chosen mixtures of nanostructured BLM with dielectric will provide impedance matching and strong light absorption in wide optical frequency range [33,34]. Then, the fact that metal interacts with light only in extremely thin skin layer will be quite favourable as only the surface electron states of BLM contribute to water splitting process. Another solution to the high reflection can be achieved in multireflection photoelectrochemical cells, where sunlight would experience multiple interactions with $MgB_2$ layers due to various optical trapping geometries (Fabry-Perot design, a semisphere, Lambertian surface, plasmonic trapping [35], etc.) which should ultimately enhance the power conversion efficiency.

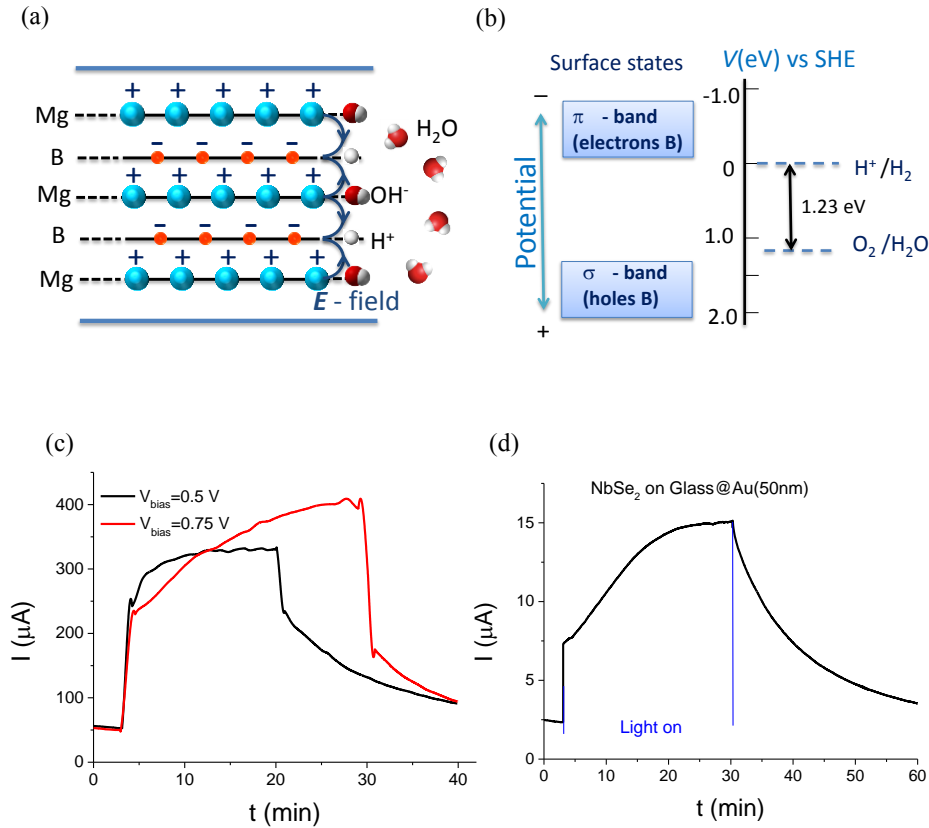

Fig. 4. (a) Schematic view of charge distribution in $MgB_2$ and electric fields responsible for ion separation. (b) Overview of the energy level positions of surface electrons in $MgB_2$ film with respect to the redox potential of $H^+/H_2$ (0 eV vs SHE). The top potential of surface electrons ($\pi$- like B) is more negative than the redox potential of $H^+/H_2$ (0 eV vs SHE) while the bottom potential of surface electrons ($\sigma$- like B) is more positive than the redox potential of $O_2/H_2O$ (1.23 eV vs SHE). (c) Photocatalytic water splitting with two electrodes made of $MgB_2$. The photocurrent as a function of time in process of water splitting with high conversion efficiency under ($V_{bias}$ =0.5 and 0.75 V). (d) Confirmation of a new class of photocatalysts material by using $NbSe_2$ layer instead of $MgB_2$ layer in photoelectrochemical cell.

It is worth mentioning that a binary BN system [36] was recently considered for the water dissociation and hydrogen production upon near-infrared light illumination. The authors have theoretically suggested a photocatalyc model in which the surface-functionalized hexagonal BN bilayers with a large-electrostatic potential difference between separated BN layers can provide the water splitting. Since pristine hBN is dielectric the authors suggested doping of BN with $H_2$ and $F_2$ to allow for electron-hole photogeneration. This fact strongly limits the power of device because of the BN is dielectric and only a small concentration of doped molecules ($H_2$, $F_2$) may contribute into photocatalyst. To the best of our knowledge, there were no successful experiments reported on this system so far.

## 4. Conclusions

In conclusion, we have proposed a new class of materials and a method for photocurrent generation in the process of catalytic water splitting under infrared-visible sunlight. The possibility to capture VS and IR light where the number of solar photons peaks is particularly appealing to the photovoltaics. Our experimental results show that both the robust nature of $MgB_2$ electronic structure and specific of charge distribution at Mg- and B- surface planes can provide the effective enhancement of photocurrent generation and hydrogen production from dissociated water molecules. We stress that BLM are commonly available and chemically stable photocatalyst which are able to generate hydrogen from water even in the absence of noble metals (Pt electrodes). There are still open questions of plasmon influence on electronic surface states in BLM, the nature of photocurrent generation dynamics, the increase of extrinsic quantum yield for metallic electrodes. However, it is clear that our results open new avenues towards the design of more energy-efficient relatively cheap photocatalytic cells where a significant fraction of sunlight energy will be used for effective hydrogen production.

## Acknowledgments

The work has been supported by EPSRC grant EP/K011022/1.